\documentclass{ws-procs975x65}

\begin{document}

\title{
Electroweak processes of the deuteron 
in effective field theory
\footnote{ 
\uppercase{T}his work is supported by 
\uppercase{N}atural 
\uppercase{S}cience and 
\uppercase{E}ngineering 
\uppercase{R}esearch 
\uppercase{C}ouncil of 
\uppercase{C}anada.}
}

\author{Shung-ichi Ando}

\address{Theory Group, TRIUMF,
4004 Wesbrook Mall, 
Vancouver, B.C. V6T 2A3, Canada\\ 
E-mail: sando@triumf.ca}

\maketitle

\abstracts{
We review our recent calculations 
of electroweak processes involving the deuteron,
based on pionless effective field theory
with dibaryon fields.
These calculations are concerned with 
neutron-neutron fusion and 
$np\to d\gamma$ at BBN energies.
}

\section{Introduction}

The study of electroweak processes 
plays an important role in few-body physics.
Effective field theory (EFT) provides
a systematic way of calculating the transition amplitudes
for those processes. 
It can also establish,
through the symmetry of QCD,
useful relations between the amplitudes for weak- and
strong-interaction processes.
Some of the important processes, 
{\it e.g.},
neutron $\beta$-decay\cite{aetal-plb04}
and the electroweak processes involving the deuteron, 
have been studied in the framework of EFT\cite{this_and_that}.
In this talk, 
we review two recent studies
on neutron-neutron fusion\cite{ak-05} and
$np\to d\gamma$ for big-bang nucleosynthesis 
(BBN)\cite{achh-05};
these studies employ pionless EFT 
with dibaryon fields (dEFT)\cite{ah-prc05}.\footnote{
We refer to it as ``dibaryon EFT'' (dEFT) in this talk.}
As regards $nn$-fusion,
we pay particular attention 
to the consequences of
uncertainties in the existing experimental data on the 
$neutron$-$neutron$ scattering length and effective range.
As for the $np\to d\gamma$ cross section
at BBN energies, 
a Markov Chain Monte Carlo (MCMC) 
is adapted to analyze the relevant experimental
data and determine the low energy constants (LECs) 
in dEFT.

\section{Neutron-Neutron Fusion, $nn\to de^-\bar{\nu}_e$}

Ultra-high-intensity neutron-beam facilities are currently 
under construction at, e.g., the Oak Ridge National Laboratory
and J-PARC and are expected to bring great progress in 
high-precession experiments concerning the fundamental properties 
of the neutron. 
Besides these experiments that focus on the properties of a single
neutron, one might consider processes that involve
the interaction of two free neutrons,
which allow the model-independent determination of the 
neutron-neutron scattering length and effective range,
$a_0^{nn}$ and $r_0^{nn}$.
In this talk, 
we first consider the $nn$-fusion process
for neutrons of very low energies 
such as the ultra-cold neutrons
and thermal neutrons.
It is worth noting that, for very low energy neutrons,
the maximum energy $E_e^{max}$ of the outgoing electrons
from $nn$-fusion
is $E_e^{max}\simeq B+\delta_N \simeq 3.52$ MeV,
where $B$ is the deuteron binding energy and 
$\delta_N = m_n-m_p$.
The value of $E_e^{max}$ is significantly larger than
the maximum energy of electrons from neutron $\beta$-decay,
$E_{e,\beta\mbox{-}decay}^{max}\simeq \delta_N\simeq 1.29$ MeV,
and thus the $nn$-fusion electrons with energies larger than 
$\delta_N$ are in principle distinguishable 
from the main background electrons of neutron $\beta$-decay.

\begin{figure}[t]
\begin{center}
\epsfxsize=7cm
\epsfbox{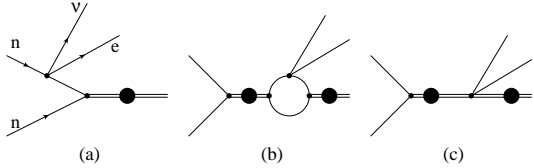}
\caption{
Diagrams for the $nn$ fusion process up to NLO in dEFT.
\label{fig;nn}}
\end{center}
\end{figure}
Diagrams for the $nn$-fusion process 
up to next-to leading order (NLO) 
are shown in Fig.~\ref{fig;nn},
from which the cross section is calculated\cite{ak-05}. 
We also include 
the Fermi function and $\alpha$-order 
radiative corrections pertaining to
the one-body interaction\cite{aetal-plb04} 
to ensure accuracy better than 1 \%
in the cross section.
The two low-energy constants (LECs),
$e_V^R$ and $l_{1A}$, 
appear in our calculation.
Using the formula for neutron $\beta$-decay\cite{aetal-plb04}
and the recent values of $G_F$, $V_{ud}$, $g_A$,
and the neutron lifetime $\tau$ in the literatures,
we deduce
$\frac{\alpha}{2\pi}e_V^R = (2.01\pm 0.40)\times 10^{-2}$.
The LEC, $l_{1A}$, which also contributes to other
processes, {\it e.g.}, $pp$-fusion and $\nu$-$d$ reactions,
can in principle
be fixed from the tritium $\beta$-decay data.
However, 
there has been no attempt to include the weak current
into the three-nucleon system in dEFT. 
So we make use of the result from the 
pionful EFT\cite{petal-prc03}, 
and obtain $l_{1A} = -0.33 \pm 0.03$.
Hence the uncertainties due to
the errors in these LECs 
and higher order terms should be
less than 1\%.
The prime uncertainty in the cross section
comes from $a_0^{nn}$ and $r_0^{nn}$,
\cite{annrnn}
\begin{equation}
a_0^{nn} = - 18.5 \pm 0.4 \mbox{[fm]} \, ,
\ \ 
r_0^{nn} = 2.80\pm 0.11 \mbox{[fm]}\, .
\label{eq;annrnn}
\end{equation}

We are now in a position 
to carry out numerical calculations
of the electron spectrum and the cross section.
Since the $nn$-fusion cross section 
obeys the $1/v$ law,
where $v$ is the relative velocity between 
the two neutrons,
we may concentrate on a particular value
of the incident neutron energy.
We consider here a head-on collision of 
two ultra-cold neutrons (UCN) 
($v_{UCN}\simeq 5$m/sec),
and thus $v=2v_{UCN}\sim 10$m/sec.
\begin{figure}[t]
\begin{center}
\epsfxsize=6cm
\epsfbox{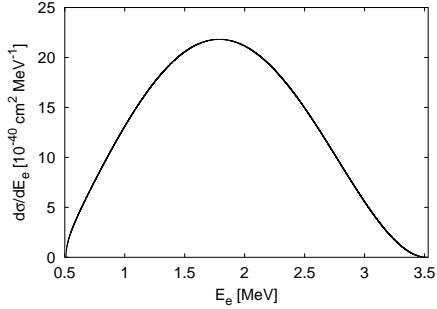}
\caption{Spectrum of the electrons from 
neutron-neutron fusion, $nn\to de\nu$.}
\label{fig;nnfusion}
\end{center}
\end{figure}
In Fig.~\ref{fig;nnfusion},
we plot the calculated electron spectrum,
$d\sigma/dE_e$, as a function of $E_e$.
As mentioned,
the electrons with $E_e>\delta_N=1.29$ MeV are 
in principle distinguishable from the electrons
coming out of neutron $\beta$-decay.
The total cross section $\sigma$ is calculated to be
\begin{equation}
\sigma = (38.6\pm 1.5)\times 10^{-40} \mbox{[cm$^2$]}\, .
\end{equation}
We find that the significant 
uncertainty ($\sim$4\%) in the cross section
comes solely from the current 
experimental errors of $a_0^{nn}$ and $r_0^{nn}$. 
Since the cross section obtained here is very small,
the experimental observation of this reaction
does not seem to belong to the near future.

\section{$np\to d\gamma$ at the BBN energies}

Primordial nucleosynthesis 
processes take place
between 1 and $10^2$ seconds
after the big bang at temperatures ranging from $T\simeq$ 1 MeV
to 70 keV. Predictions of primordial light element abundances, D,
${}^3$He, ${}^4$He, and ${}^7$Li, and 
the comparison of them
with observations are a crucial test of the standard 
big bang cosmology.
The uncertainties in these predictions are 
dominated by the nuclear physics input for the reaction
cross sections.
Reaction databases are continuously updated\cite{bbn},
with more attention now paid to the error budget.
The cross section of the $np\to d\gamma$ process at
the BBN energies has been thoroughly 
studied by using pionless
EFT up to 
N$^3$LO by Chen and Savage\cite{cs-prc99}, 
and up to N$^4$LO by Rupak\cite{r-npa00}.
In this part of talk, 
we present
an estimation of the cross section 
employing a new method, {\it i.e.}, a combination of
dEFT up to NLO and an MCMC analysis with the aid of
the relevant experimental data.
We find that this method leads to a 
result comparable with that obtained by Rupak,
and we discuss that
the estimated $np\to d\gamma$ cross section  
at the BBN energies is reliable to within 1\%. 

\begin{figure}[b]
\begin{center}
\epsfxsize=7cm
\epsfbox{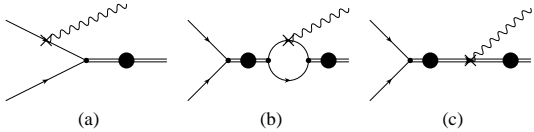}
\caption{Diagrams for the $np\to d\gamma$ process up to NLO
in dEFT.
\label{fig;npdg}}
\end{center}
\end{figure}
Diagrams for the $np\to d\gamma$ process up to NLO 
in dEFT are shown in Fig.~\ref{fig;npdg}.
From these diagrams we calculate the amplitudes
for the $S$(${}^1S_0$ and ${}^3S_1$)-
and $P$-waves of the initial two-nucleon.  
We note that since the ${}^3S_1$ amplitude 
is highly suppressed due to the orthogonality of the scattering
and bound ${}^3S_1$ states, we neglect it in our calculations.
Using these amplitudes, we can easily calculate
the cross section for $np\to d\gamma$.

\begin{table}[t]
\tbl{Values of parameters
\label{tab;parameters}
}
{\footnotesize
\begin{tabular}{c|cc} 
\hline
         & MCMC & Prev. Method \\ \hline
$a_0$    & $-23.7426\pm 0.0081$ & $-23.749\pm 0.008$ \\ 
$r_0$    & $2.783\pm 0.043$     & $2.81\pm0.05$ \\
$\rho_d$ & $1.7460\pm 0.0072$   & $1.760\pm0.005$ \\
$l_1$    & $0.798\pm 0.029$     & $0.782\pm 0.022$ \\ 
\hline
\end{tabular} }
\end{table}
Five parameters,
$a_0$, $r_0$, $\gamma$, $\rho_d$, and $l_1$, 
appear in the amplitudes. 
We determine the values of the four parameters,
$a_0$, and $r_0$, $\rho_d$, and $l_1$,
by the MCMC analysis 
of the relevant low energy
experimental data; 
the total cross section of the $np$ scattering at the energies 
$\le$5 MeV (2124 data)
from the NN-OnLine web page, 
the $np\to d\gamma$ cross section from 
Suzuki {\it et al.}\cite{suzuki_etal}
and Nagai {\it et al.}\cite{nagai_etal}
including two thermal capture data\cite{cox_etal},
the $d\gamma\to np$ cross section from Hara {\it et al.}\cite{hara_etal}
and Moreh {\it et al.}\cite{moreh_etal},
and the photon analyzing power 
from Tornow {\it et al.}\cite{tornow_etal} 
and Schreiber {\it et al.}\cite{schreiber_etal}.
Meanwhile, we constrain $\gamma$ 
from the accurate value of $B$. 
In Table~\ref{tab;parameters}
we give our estimates of the parameters 
obtained from the present MCMC analysis
along with the values obtained in the previous method 
(``Prev. Method'').\footnote{
The values of the effective ranges, $a_0$, $r_0$, and $\rho_d$,
are taken from Ref.\cite{kn-zpa75},
and the value of $l_1$ is obtained from the averaged value
of the two thermal capture rates\cite{cox_etal}.
}
We find small differences ($\le$2\%) between 
the values of the parameters for the two
cases; we will come back to this later.

In Table~\ref{table;results}
the theoretical estimates of the $np\to d\gamma$
cross section at BBN energies are given as
a function of the initial two-nucleon energy $E$
in the center of mass (CM) frame.
The column labeled ``dEFT(MCMC)'' 
gives our preliminary results
for the mean values and standard deviations 
obtained in MCMC.
Table~\ref{table;results} also shows
the results of four other methods:
``dEFT(Prev. Meth.)''
based on the parameter set ``Prev. Method'' 
in Table~\ref{tab;parameters},  
pionless EFT up to N${}^4$LO by Rupak,
a high-precision potential model calculation including the
meson-exchange current by Nakamura,
and an R-matrix analysis by Hale.
Good agreement is found among the different approaches
except that the results of ``dEFT(Prev. Meth.)'' 
at $E=0.5$ and 1 MeV and those of Hale exhibit
some deviations, which are $\sim$0.6\% 
in the former and go up to 4.5 \%
in the latter.
The $\sim$0.6\% difference at $E=0.5$ and 1 MeV 
between ``dEFT(MCMC)'' and ``dEFT(Prev. Meth.)'' 
is significant compared to the small $\sim$0.3\% statistical 
errors obtained here. 
This difference can be accounted for 
by higher order terms that are not included in 
the amplitudes of dEFT up to NLO.
By including the higher order terms 
associated with the $P$-wave scattering volumes\cite{cs-prc99},
we can reproduce the ``dEFT(MCMC)''
results at $E=0.5$ and 1 MeV
in ``dEFT(Prev. Meth.)''.
This implies that the values fitted by MCMC
mimic the roles of the higher order terms. 
Since our results ``dEFT(MCMC)''  
agree quite well with those of 
Rupak and Nakamura's calculations,
and since
in the N$^4$LO pionless EFT calculation by Rupak,
various corrections due to the higher order terms
have been studied, we infer that 
the estimated $np\to d\gamma$ cross section 
at the BBN energies
should be reliable within 1\% accuracy.
A dEFT calculation provides a systematic 
perturbation scheme
and a simple model-independent expression 
for the amplitudes
in terms of a finite number of LECs.
As demonstrated above, 
the combination of a dEFT calculation 
and an MCMC analysis
of available experimental data
would be a useful method to deduce
reliable cross sections for 
other few-body processes. 

\begin{table}[t]
\tbl{Theoretical estimates of the $np\to d\gamma$
cross sections at the BBN energies.
$E$ is the initial two-nucleon energy in CM frame.
See the text for more details.
\label{table;results}}
{\footnotesize
\begin{tabular}{c|ccccc} \hline
E(MeV)               & dEFT(MCMC) & dEFT(Prev. Meth.) & 
 Rupak & Nakamura & Hale \\ \hline
$1.265\times10^{-8}$ & 333.8(4) & 333.7(15) & 
 334.2(0) & 335.0 & 332.6(7)  \\ 
$5\times10^{-4}$ & 1.667(2) & 1.666(8) & 1.668(0) & 1.674 & 1.661(7) \\
$1\times10^{-3}$ & 1.171(1) & 1.171(5) & 1.172(0) & 1.176 & 1.167(2) \\
$5\times10^{-3}$ & 0.4979(6) & 0.4976(21) & 0.4982(0) & 0.4999 & 0.4953(11) \\
$1\times10^{-2}$ & 0.3321(4) & 0.3319(14) & 0.3324(0) & 0.3335 & 0.3298(9) \\ 
$5\times10^{-2}$ & 0.1079(1) & 0.1079(4) & 0.1081(0) & 0.1084 & 0.1052(9) \\
0.100            & 0.06341(7) & 0.0634(2) & 0.06352(5) & 0.06366 & 0.0605(10)\\
0.500            & 0.03413(8) & 0.0343(1) & 0.0341(2) & 0.03416 & 0.0338(8) \\
1.00             & 0.03502(10) & 0.0352(2) & 0.0349(3) & 0.03495 & 0.0365(8)\\
\hline
\end{tabular}
}
\end{table}

The author would like to thank K. Kubodera, 
R.~H. Cyburt, S.~W. Hong, and C.~H. Hyun 
for collaboration.

\end{document}